\documentclass[prb,preprint]{revtex4-1} 

\usepackage{xcolor}
\usepackage{amsmath}  
\usepackage{amsfonts} 
\usepackage{graphicx} 

\usepackage{hyperref}

\hypersetup{
    urlcolor=cyan,
    pdftitle={Monte Carlo, blocking,  and  inference:  How to measure the renormalization group flow},
    pdfpagemode=FullScreen,
    }

\newcommand{\ham}{\mathcal{H}}
\newcommand{\kk}{K}
\newcommand{\kkk}{\boldsymbol{{K}}}

\newcommand{\bn}{\boldsymbol{ n}}
\newcommand{\bmm}{\boldsymbol{ m}}
\newcommand{\bs}{\boldsymbol{ \sigma}}
\newcommand{\mean}[1]{\left \langle #1 \right \rangle}

\begin{document}

\title{Monte Carlo, blocking,  and  inference: \\ How to measure the renormalization group flow}

\author{Luca Di Carlo}

\affiliation{Joseph Henry Laboratories Department of  Physics, Princeton University, Princeton, New Jersey 08544, USA }

\date{\today}

\begin{abstract}
Renormalization group theory is a powerful and intriguing technique with a wide range of applications. 
One of the main successes of renormalization group  theory is the  description of  continuous phase transitions and the development of  scaling theory. 
Most  courses  on phase transitions focus on scaling and critical exponents, while less attention is paid to universality,   renormalization group flow, and the existence of a unique fixed point, which are the ultimate reasons why  scaling theory is so effective in describing continuous phase transitions.
We use a combination of Monte Carlo simulations and  real space renormalization group theory to  determine the renormalization group flow    and  to show the existence of a universal fixed point in the context of the ferromagnetic Ising model.
\end{abstract}

\maketitle
\section{Introduction} 

Continuous phase transitions occur when a material undergoes a spontaneous change in its symmetry or its state. A paradigmatic example of a continuous phase transition is the ferromagnetic transition in which a magnetic material loses its spontaneous magnetization  as it is heated beyond the critical temperature.  The hallmark of continuous phase transitions is the divergence of the correlation length. 
How the singular behavior of  various thermodynamic quantities, as characterized by critical exponents, is   related to the divergence of the correlation length is well described by  scaling theory.\cite{Widom}

A particularly intriguing experimental observation is the existence of shared critical properties across different physical systems, characterized by identical  critical exponents. For example,  different ferromagnetic materials composed of different atoms or possessing distinct crystalline structures exhibit similar behavior close to the critical temperature. The range of applicability of  universality  is not  restricted  to ferromagnetic systems and  extends to all continuous phase transitions with the same symmetries. Critical opalescence of a liquid close to the critical point is described by the same critical exponents as uniaxial ferromagnets, and the superfluid phase transition shares common features with some ferromagnetic transitions.\cite{Kardar,Cardy} For this reason continuous phase transitions can be organized into universality classes sharing the same critical exponents. 

Close to the transition point the only relevant length scale is the correlation length $\xi$, and its divergence is solely responsible for the divergencies of  various thermodynamic quantities. In this sense,  the microscopic details become unimportant in determining the properties of a system close to its critical point, which eventually led to the development of  renormalization group theory. In a nutshell,  renormalization group theory consists of systematically removing small scale degrees of freedom and observing how their removal affects the large scale physics. This approach can alternatively be interpreted as probing the system at different length scales. The renormalization group flow  tracks how the effective theory describing the system changes at different length scales. The aim of this paper is to numerically explore some of these ideas in the context of the Ising model. 

The Ising model, which is one of the simplest models exhibiting a ferromagnetic transition,  has been extensively used to study the physics of continuous phase transitions. The Ising model is typically taught to test  scaling theory and to determine the critical exponents. We will   use the Ising model to study numerically the theoretical foundations of  scaling theory and universality in continuous phase transitions, namely the renormalization group.  By using a combination of blocking and inverse Ising techniques, we will determine numerically the renormalization group flow  and test one of the most intriguing predictions of renormalization group theory: the existence of a fixed point.

We assume that the reader is familiar with the Ising model, with the Markov-chain Monte Carlo technique,\cite{MCMC} and with some of the basics of renormalization group theory. The paper is organized as follows. In Sec.~\ref{BasicIsing} we introduce the Ising model, and in Sec.~\ref{generalRG} we review some of the basic ideas of  renormalization group theory such as  the critical manifold, renormalization group flow, and the fixed point. For a clear and accessible treatment of  renormalization group theory the reader may refer to the book by Cardy,\cite{Cardy} in particular, Chaps.~1 and 3. In Sec.~\ref{IIP}  we introduce some necessary numerical methods, in particular, a method to efficiently solve the Inverse Ising problem. Finally, in Sec.~\ref{numerical} we describe a procedure to measure the renormalization group flow and discuss some numerical results.

\section{The Ising model} 
\label{BasicIsing} 
The Ising model is a minimal model capable of capturing the fundamental aspects of continuous phase transitions.
In its original formulation the Ising model describes a system of binary variables, called spins,  on a two-dimensional (2D) lattice  with  nearest-neighbor ferromagnetic interactions. We denote  the spins as $\{\sigma_{\boldsymbol n } \}$, where $\boldsymbol n = (n_x, n_y)$, $n_{x,y} = 1,\dots L$,  and $L$ is the linear size of the system. The  Boltzmann probability distribution defines the statistical properties of the Ising model, 
\begin{equation} 
\mathcal  P [ \bs] =  \frac 1 Z e^{- \ham[ \bs]/k_B T },  \qquad  \mathcal \ham[\bs] = - \frac J 2 \sum_{|\bn- \bmm| =1} \sigma_{\bn}  \sigma_{\bmm},
\label{BasicIsingP}
\end{equation} 
where the sum $\sum_{|\bn - \bmm| = 1} $ is  over all pairs of nearest-neighbor spins and the coupling constant $J>0$. The function $\mathcal H$ plays the role of an energy function, and we will improperly call $\mathcal H$ the Hamiltonian or the energy.  

Depending on the value of the  ratio $\kk = J/k_B T$, the Ising model can be in two distinct phases, which are determined by the magnetization $m$ of the system, 
\begin{equation} 
	m = \frac 1 {L^2} \sum_{\bn} \mean{ \sigma _{\bn}}  , 
	\label{magno}
\end{equation}
where the angular brackets indicate the average over the probability distribution in Eq.~\eqref{BasicIsingP}.\cite{MCMC_is_hard} The paramagnetic phase ($ m\simeq 0$) and the ferromagnetic phase ($ m \simeq \pm 1)$ are separated by a critical point at which the correlation length $\xi$ diverges.  The correlation length is the decay rate of the spatial correlation function,
\begin{equation} 
	C(\bn,\bmm) =  \mean{\sigma_{\bn} \sigma_{ \bmm} } -  \mean{\sigma_{\bn}} \mean{\sigma_{\bmm}} .
	\label{aioli} 
\end{equation} 
Translational symmetry constrains the correlation function 
 to depend only on the  distance $|\bn - \boldsymbol{m}| = |n_x - m_x| + |n_y - m_y| $. Therefore, 
\begin{align} 
	& C(r) =\frac 1 N  \frac{1}{ 4 r }\sum_{|\bn - \bmm | =   r } \mean{\sigma_{\bn} \sigma_{ \bmm} } - \mean{\sigma_{\bn}}  \quad \qquad (r>0) , \\
	& C(0) =\frac 1 N \sum_{\boldsymbol n  } 1 -  \mean{\sigma_{\boldsymbol n} }^2 , 
\end{align}
where  $r = |\bn - \boldsymbol{m}| = |n_x - m_x| + |n_y - m_y|$.
The factor of $4r$ accounts for the number of spins at a distance $r$ from a given site at fixed value of $\boldsymbol n$ in two dimensions.

\begin{figure}[!]
\centering
\includegraphics[width = 0.8\textwidth]{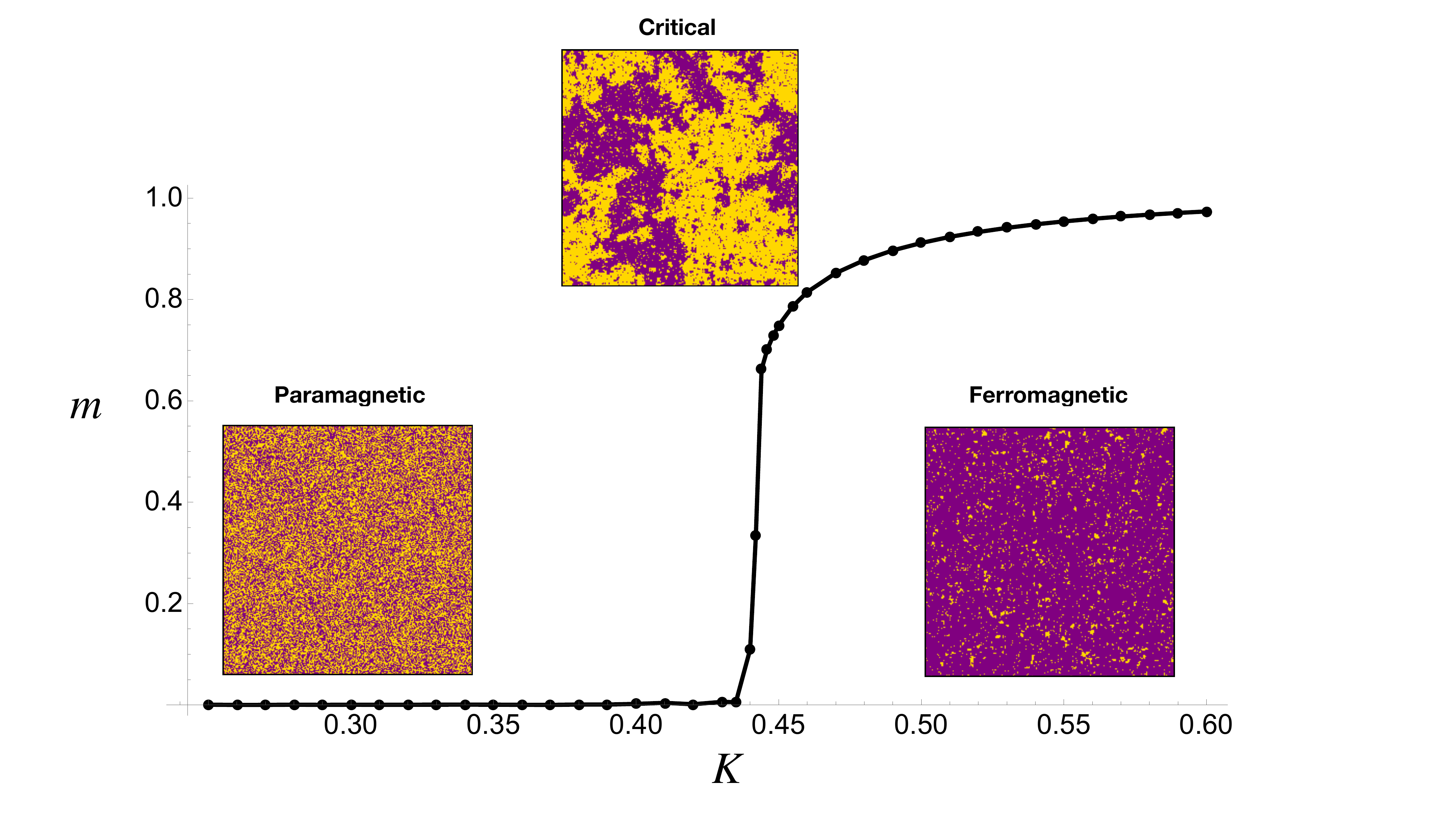}
\caption{(Color Online)
The magnetization $m$ as a function of the coupling $\kk$ for the 2D Ising model  for $L = 128$ and periodic boundary conditions. The transition between the disordered phase, $m \simeq 0$ for small $\kk$, and the ordered phase, $m\simeq 1$ at large $\kk$, occurs at $\kk_c \simeq 0.44$.  Also shown is some typical configurations in the disordered and ordered phases and near the  critical point. 
The configurations are obtained using the Metropolis--Hasting algorithm\cite{MCMC2} (see the IsingRG.py file in Ref.~\onlinecite{suppl}).}
\label{appropriateLable}
\end{figure}

\begin{figure}[h]
\centering
\includegraphics[width = 0.9\textwidth]{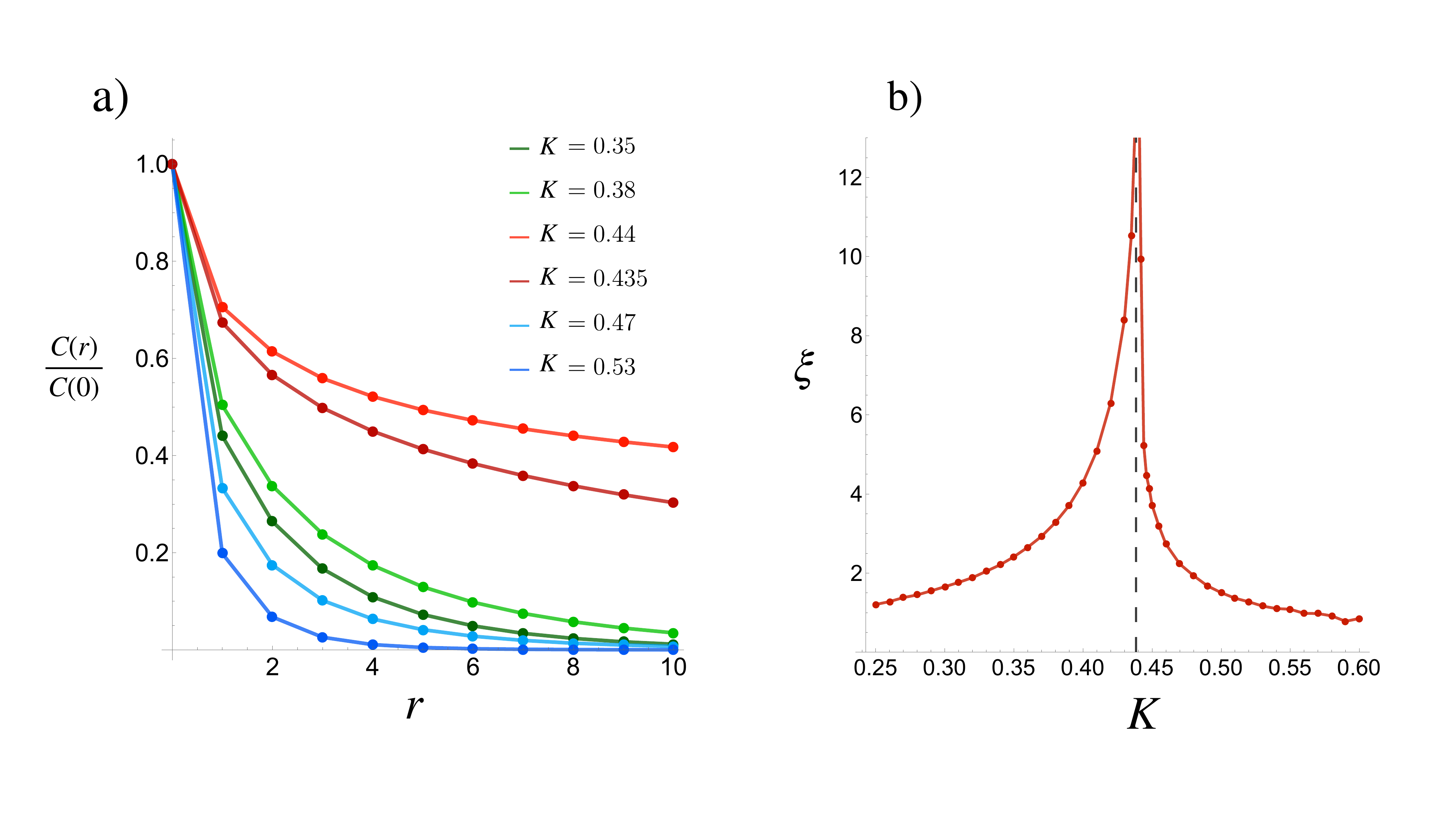}
\caption{
(a) The correlation function $C(r)$ exhibits a rapid decay in the paramagnetic and ferromagnetic phases, as represented by the green and blue curves respectively. Near the critical point (red curves)  the decay of   $C(r)$ is notably slower.  For convenience, we  show the normalized quantity $C(r)/C(0)$. 
(b) The correlation length is  estimated by fitting the exponentially decaying function $e^{-r/\xi}$ to the correlation function. The correlation length  $\xi$ diverges at the transition point.}
\label{appropriateLable2}
\end{figure}

In Fig.~\ref{appropriateLable} we show how the average magnetization $m$ in Eq.~\eqref{magno} depends on the coupling constant $\kk$. In Fig.~\ref{appropriateLable2} we show how the correlation function $C(r)$ and the correlation length $\xi$ vary with     $\kk$. 

\medskip \noindent \textit{Problem 1: Monte Carlo simulation  of the 2D Ising model}. Use the Metropolis--Hasting algorithm\cite{MCMC}  and $N_s$  samples $\{ \sigma_{\bn} (t)\}_{t= 1, \dots,N_s}$ from the probability distribution in Eq.~\eqref{BasicIsingP} with  $\kk = 0.42$ and $L = 128$. 
Measure the average magnetization $m$ and the correlation function $C(r)$. Fit the latter to an  exponential to estimate the correlation length. Repeat the same measurements for different values of $\kk$ in the range $\kk \in [0.3,0.6]$ and determine how   $m$ and    $\xi$ depend on $\kk$. The results should be consistent with the ones shown in Figs.~\ref{appropriateLable} and \ref{appropriateLable2}.

\section{Renormalization group theory} 
\label{generalRG}

\begin{figure}[h]
\includegraphics[width = \textwidth]	{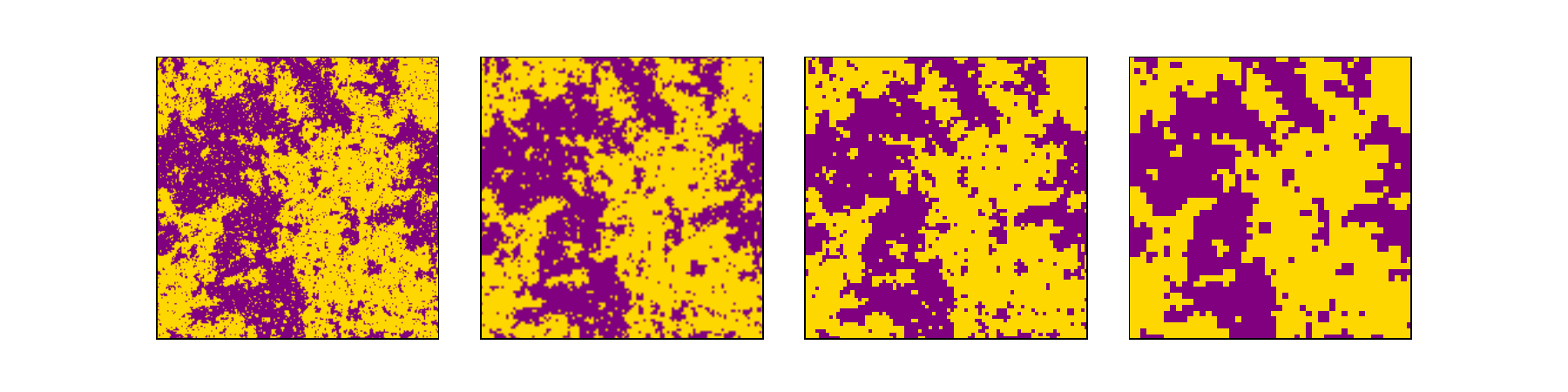}
\caption{Effect of the Kadanoff blocking on a equilibrium configuration of the Ising model. The microscopic details are washed away when increasing the block size $b$. The reader should be able to reproduce these figures by following the ``Simulating the Ising model" section of the MeasureRGflow.ipynb python notebook in Ref.~\onlinecite{suppl}. The effect of the Kadanoff blocking transformation on the Ising model can be better visualized using  the  program IsingBlocking.pde  in Ref.~\onlinecite{suppl}. 
}
\label{CoorsLight} 
\end{figure}

The conceptual breakthrough that led to a deep theoretical understanding of    continuous phase transition was  the intuitive idea that in  correlated systems with long-range correlations, that is,  a divergent correlation length,  the  details of the microscopic interactions are unimportant when it comes to macroscopic/large-scale properties. If we restrict ourselves to  Ising-like models, we can make this statement more concrete by considering the  Kadanoff blocking transformation.\cite{Koff0,Koff1,Barbie}  The Kadanoff blocking transformation involves  two steps: coarse graining and rescaling. The first step  consists of grouping neighboring spins, within a block of linear dimension $b$,  and defining new block variables $\boldsymbol \sigma^{(b)}$ to obtain a coarser description of the system in which small-scale fluctuations are washed away. To be  specific, we group spins into $b\times b$ blocks $\mathcal B_i$, each containing $b^2$ spins, and assign to each of these new block variables a value $\sigma^{(b)} = \pm 1$ according to the majority rule:
\begin{align}
 \sigma^{(b)}_i & = \mathcal M \left(\sum_{j \in \mathcal B_i} \sigma_j\right)  \label{Keglevich}
 \\ 
 \mathcal M(\sigma) & =
 \begin{cases}
    +1   &  (\sigma>0) \\ 
   -1 & (\sigma<0) \\ 
   \mbox{rnd}(\pm 1) & (\sigma=0) \\  \label{Keglevich}
\end{cases}      
\end{align}
If the majority of the spins inside the block are $+1$, we set the value of the block variable to $1$; if the majority are $-1$, we set the value of the block variable to $-1$.  If $b$ is even and $\sigma = 0$,  we randomly choose whether the block variable is $\pm 1$.  We will often refer to this blocking transformation as coarse-graining, and we will refer to the blocked variables as the coarse-grained degrees of freedom. In Fig.~\ref{CoorsLight}  we show the affect of the Kadanoff blocking transformation on an equilibrium configuration of the ferromagnetic Ising model for different values of the block size $b$.
 
The second step of the Kadanoff blocking transformation is rescaling. The distance between two  neighboring spins in the original lattice is exactly one lattice spacing,  but after performing a coarse graining with block size b, the distance between two neighboring block variables is now $b$ lattice spacings. Crucially, if we want to compare the original system with the coarse grained one, we have to rescale the coarse grained system by a factor of $b$. In this way the distance between neighboring block variables remains one lattice spacing for any value of the block size $b$.

The statistical properties of the coarse-grained configurations are different from those of the original system; measuring the same observable at different levels of coarse graining yields different results.\cite{MCMC2,Cardy} If the original system is described by a Hamiltonian $\ham [\bs]$, the effective Hamiltonian $\ham_b[\bs^b]$ describing the coarse-grained degrees of freedom is in general different at each level of coarse-graining. How the Hamiltonian changes under coarse-graining is   called the renormalization group flow.

\subsection{Critical manifold, renormalization group fixed point, and universality}
The renormalization group flow, which is how the Hamiltonian changes at different levels of coarse-graining, is the key to understanding universality and the effectiveness of the scaling theory.  The qualitative behavior of the renormalization group flow  depends on how close the system is to the critical point. We can distinguish three qualitatively different scenarios depending on whether the system is in the paramagnetic or ferromagnetic phase or at the critical point. 	

In the paramagnetic phase the  magnetization  is zero and the spins are weakly correlated. 
As the coarse-graining process progresses, the system will converge toward block variables that are  random and uncorrelated. Therefore, the Hamiltonian of the system flows toward a completely random model, which can be interpreted as either at an infinite temperature  or a zero coupling fixed point. In the ferromagnetic phase,  coarse-graining leads to block variables that are more and more ordered, and thus the Hamiltonian flows to a strongly coupled model, which may  be interpreted as a zero temperature fixed point. Both these scenarios are trivial (zero or infinity) and not very exciting.  The renormalization group flow becomes more interesting close to the critical point, which divides the paramagnetic and the ferromagnetic phases. In this case the renormalization group flow will approach neither of the trivial fixed points, but will converge to the nontrivial finite fixed point.

\begin{figure}[h]
\centering
\includegraphics[width = 0.99\textwidth]{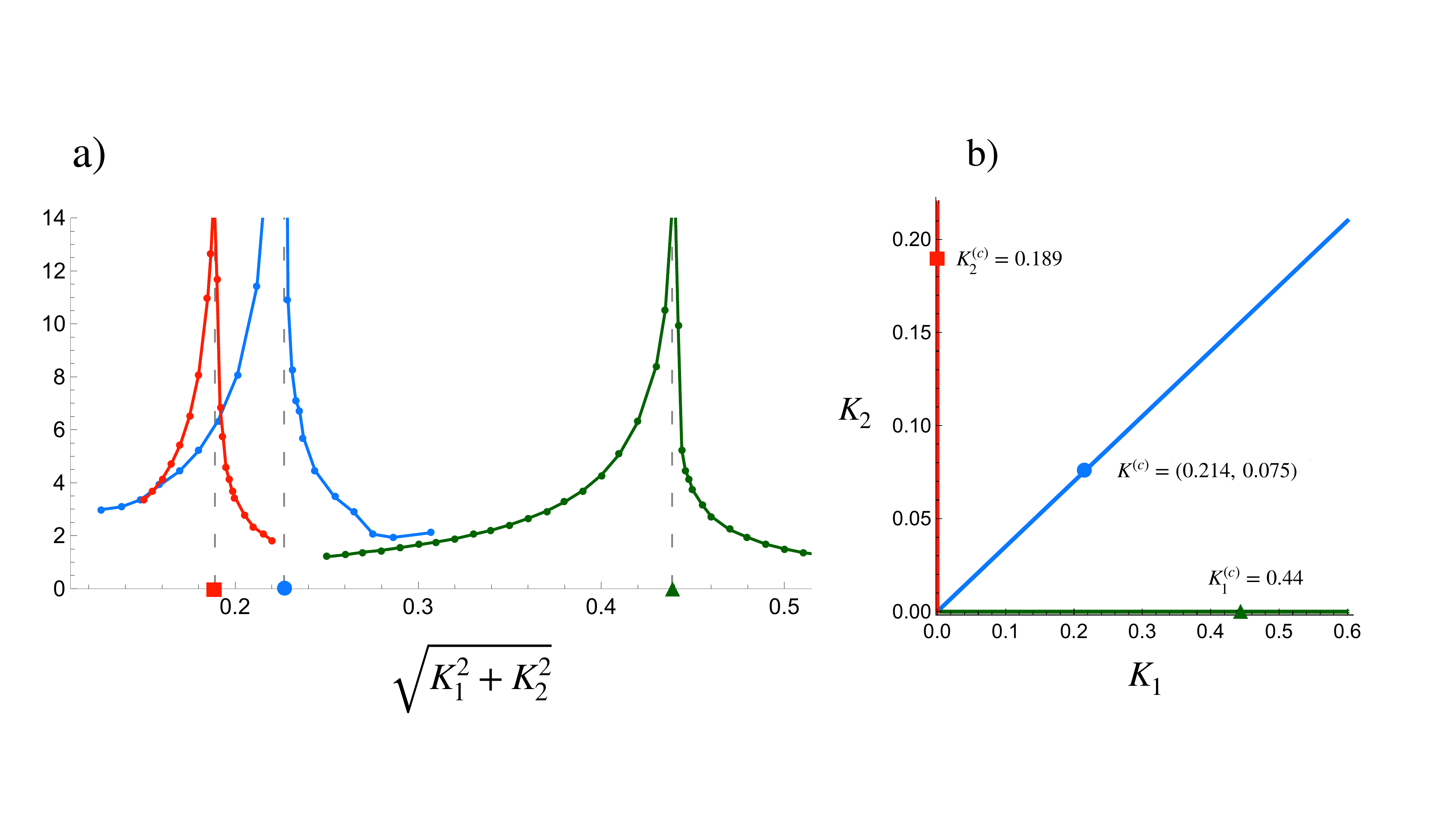}
\caption{(Color online) The correlation length   for a  $L = 128$ Ising model with periodic boundary conditions. We scan the parameter space in three  directions  and measure the correlation length $\xi$.  (a) The red, green, and blue curves correspond to scanning the  $(\kk_1, \, \kk_2) $ plane on the $\kk_2$ axis, $\kk_1$ axis, and on the $\kk_2 = 0.35 \times \kk_1$ line as shown in (b), respectively . All  three cases  cross a critical point with a divergent correlation length. 
}
\label{amazza}
\end{figure}

We now consider an example of  an Ising model with nearest-neighbor $\kk_1$ and next-nearest neighbor $\kk_2$ interactions. If both $\kk_1$ and $\kk_2$ are very small, the system is in the paramagnetic phase. Conversely, if the couplings are sufficiently large, the system is in the ferromagnetic phase. We can think of doing three  numerical experiments, which are summarized in Fig.~\ref{amazza}. The first experiment consists of chosing $\kk_2 = 0$ and gradually increasing     $\kk_1$ (green curve);  eventually $\kk_1$ will be large enough to cross a critical point of the type 
\begin{equation} 
\kk_1 = \kk_1^c \qquad   \kk_2 = 0 , 
  \label{critical1}
\end{equation} 
and enter the ferromagnetic phase.
For the second numerical experiment we keep $\kk_1=0$ and  gradually increase $\kk_2$ (blue curve). As in the previous case, we will eventually enter the ferromagnetic phase after crossing a critical point of the type, 
\begin{equation}
\kk_1 = 0   \qquad  \kk_2 = \kk_2^c .
\label{critical2}
\end{equation} 
Finally, we can start with small $\kk_1$ and $\kk_2$ and increase both of them gradually. In this case we also expect to cross a critical point   with both $\kk_1$ and $\kk_2$ different from zero.

Note that there is no unique critical point, in the sense that the values of the couplings corresponding to a system with infinite correlation length are not unique.   We can  convince ourselves that there are infinitely many possible combinations of the parameters $\kk_1,\kk_2$ for which the system is at the critical point. We call the \textit{critical manifold} the set of values of the couplings $\kk_1,\kk_2$ for which the correlation length $\xi(\kk_1,\kk_2)$ is infinite. 
The prediction of  renormalization group theory is that the renormalization group flow converges to the same universal fixed point independently of the specific values of $\kk_1$ and $\kk_2$ if they are on the critical manifold $\xi(\kk_1, \kk_2) = \infty$.

\begin{figure}[h]
 	\includegraphics[width = \textwidth]{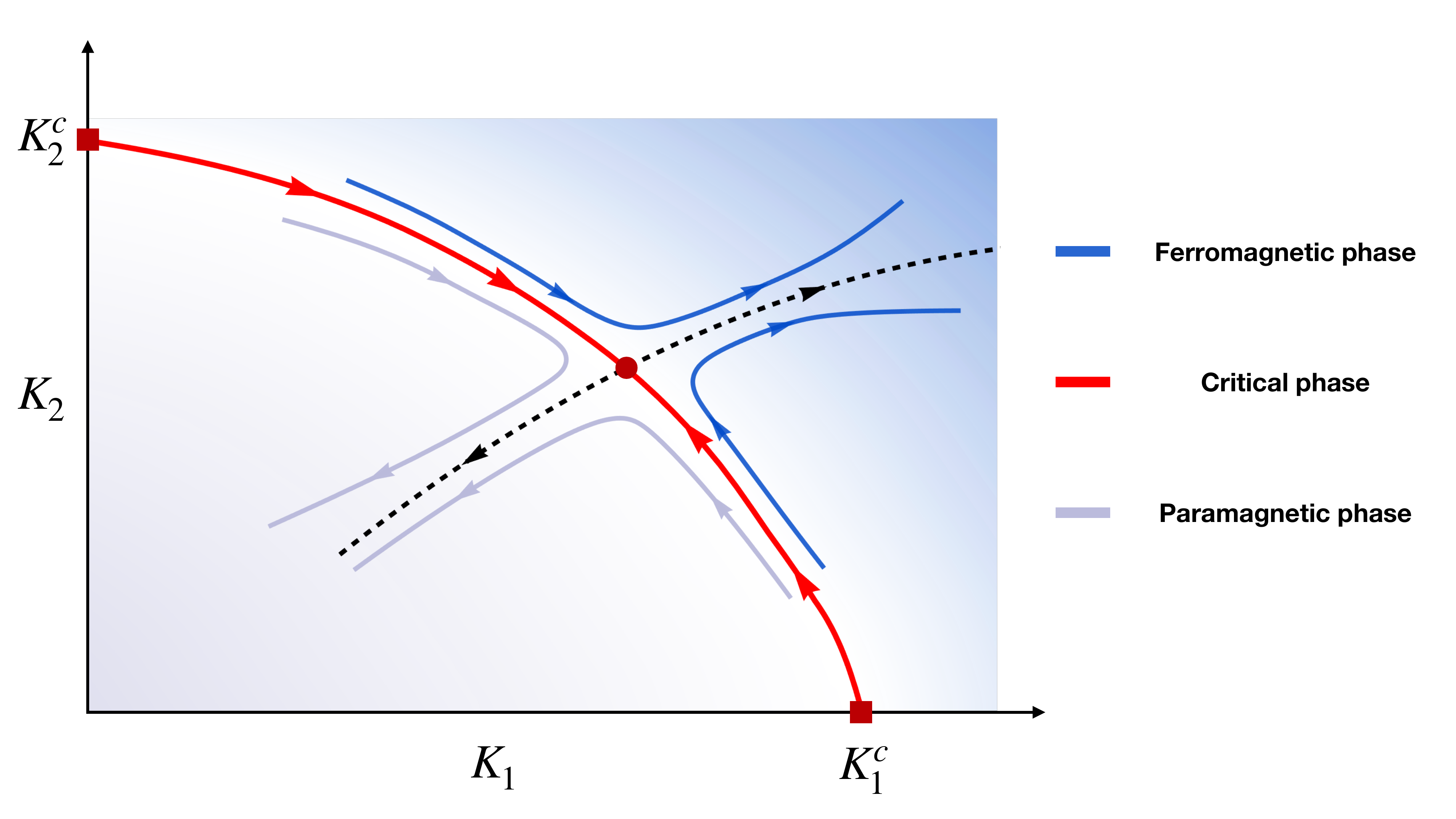}
	\caption{(Color online) Qualitative sketch of the renormalization group flow of a 2D Ising model with nearest-neighbor $ K_1$ and next-nearest-neighbor $ K_2$ interactions.  The arrows show how the model's parameters evolve under coarse-graining. The qualitative behavior of the renormalization group flow depends on the value of the parameters $(K_1,  K_2)$. When the couplings are large (blue region), the system is in a ferromagnetic state and the renormalization group flow (blue curves) flows to a strongly interacting model $\kkk  \to \infty$.  When the couplings are small, (gray region) the system is in the paramagnetic phase, and the renormalization group flow (gray curves) drives the system to a non-interacting $ \kkk = 0$ model.  In the critical regime, or more precisely, on the critical manifold $\xi(K_1, K_2) = \infty$, the renormalization group flow converges to a universal fixed point (red dot)  which is independent of the particular value of the pair  $(K_1,\,   K_2)$. For example, the two systems with coupling $(K_1^c, 0)$ and  $(0, K_2 ^c$) both converge to the same fixed point. }
    \label{seborrea}
\end{figure}

This behavior can be summarized in the renormalization group flow diagram shown in Fig.~\ref{seborrea}, which describes how the parameters of the model change under a coarse-graining transformation. The general picture we obtain from Fig.~\ref{seborrea} is that in the ferromagnetic  and   paramagnetic phases the renormalization group flow converges to trivial fixed points, which are the strong coupling $\kkk  = \infty$ and zero coupling regime $\kkk = 0$, respectively. Conversely, on the critical manifold (red curve) where the correlation length $\xi$   diverges, the renormalization group flow converges to a unique fixed point (red dot), which is independent of the specific value of the model's parameters before applying the coarse-graining transformation.\cite{Escaping}  For this   reason  different systems display the same behavior close to a critical phase transition, because their large scale physics is described by the same fixed point theory.

We can think of this Kadanoff blocking procedure as a thought experiment or a theoretical tool to study critical phase transitions. If we take this picture seriously, we should be able to actually measure the renormalization group flow. Although the scaling theory predicted by renormalization group theory can be easily tested numerically, it is not immediately obvious how to measure the renormalization group flow and show   that all systems on the critical manifold converge to the same fixed point.

To measure  how various observables scale under coarse-graining is straightforward. We just have to measure the same observables at different levels of coarse-graining.\cite{MCMC2} It is less clear how to measure how the model's parameters, or  the effective Hamiltonian, change under coarse-graining.

To tackle  this problem in a slightly different context,  Shenker and Tobochnik\cite{MCMC_RG} used an ingenious strategy using a Monte Carlo renormalization group approach which involves comparing the effect of the blocking transformation on systems of different sizes and parameters. In this way they  estimated the beta-functions of the model. The beta functions describe how the effective description of the system changes with the length scale at which we probe the system, and are intimately connected to the renormalization group flow.\cite{gold} Remarkably,  this approach made possible extracting information on the structure of the renormalization group flow in a time with   very limited computational resources. 
We will use a different approach, which involves solving the  inverse Ising problem. 

\medskip \noindent \textit{Problem 2. Magnetization distribution and blocking transformation}. (a) From the Ising model configurations  of Problem 1  determine the histogram of the  magnetization  $m$  for various values of the coupling $K$.  Discuss how the histogram of the magnetization differs in the paramagnetic and ferromagnetic phases and near the critical point. (b) Apply the Kadanoff blocking transformation  to the same configurations  with $b = 2$, and determine the histogram of the distribution of the magnetization of the blocked system,
$m =  \sum_{\bn} \sigma_{\bn}^{(b)}$.
Repeat this procedure for  $b = 2,3,4,5$, and discuss how the effect of the Kadanoff blocking transformation differs in the three regions. Is this behavior consistent with the theoretical prediction of Fig.~\ref{seborrea}?

\medskip \noindent \textit{ Problem 3: Ising model with next-nearest neighbor interactions}. 
Simulate  a $L = 120$  Ising model with  $ K_1 = 0.16 $ and  $ K_2 = 0.04$
\begin{equation}
	\mathcal P (\bs) = \frac 1 Z \exp\left(\frac 12  K _1  \sum_{|\bn - \boldsymbol{m}| = 1 }\sigma_{\bn} \sigma_{\boldsymbol{m}} +
	\frac 12  K _2  \sum_{|\bn - \boldsymbol{m}| = 2}\sigma_{\bn} \sigma_{\boldsymbol{m}}
	 \right), \label{configurations}
\end{equation}
and measure the magnetization and the correlation length. Increase   $ K_1$ and $ K_2$ on the line $ K_1/ K_2 = 4$ and determine how  $m$ and  $\xi$ change.

\section{ The Inverse Ising problem} 
\label{IIP} 
Given a spin Hamiltonian $\mathcal H[\bs]$  and $\beta \equiv 1/k_BT$, we can sample equilibrium configurations  of the probability distribution $\mathcal{P} = e^{-\beta \ham[\bs]}/Z$ using a Markov-chain Monte Carlo algorithm,\cite{MCMC2} and compute its moments  and correlations such as, 
\begin{equation} 
\mu_{\bn} = \mean{ \sigma_{\bn}}  \qquad  C_{\bn \bmm} = \mean{\sigma_{\bn}\sigma_{\bmm}}.
\label{GoatMan}
\end{equation} 
Sampling from the probability distribution is what we call the direct Ising problem. In the inverse Ising problem the quantities in Eq.~(\ref{GoatMan}) or  some equilibrium samples  of an Ising model $\{\sigma_{\boldsymbol {n}}(t) \}$ are given and one computes the Hamiltonian $\ham$ that would generate these moments and correlations.  In particular, given $N_s$ {independent} samples of the form,
\begin{equation}
	\{\sigma_{\bn} (t) \}_{n = 1, \dots, L^2 } ^{t= 1, \dots, N_s},
	\label{datta}
\end{equation}
where $\bn$ labels the position on the lattice and $t$  labels the samples, we wish  to find the probability distribution $\mathcal P^\star \propto e^{-\ham^\star}$, or better the Hamiltonian $\ham^\star$, that most likely generated the data.  Here we have incorporated $\beta$ into $\ham^\star$.

\subsection{Maximum likelihood} 

One way of finding the Hamiltonian $\ham^\star$ that best fits the data is to use   the maximum likelihood approach. This approach consists of finding the probability distribution $\mathcal P_{\mathrm{ML}}$ that most likely  generated the data. 
One difficulty is that the space of possible probability distributions of $L^2$ binary variables is quite large ($2^{L^2}-1$ degrees of freedom,   corresponding to  every possible state of the system). A possible way out is to make some assumptions for the form of   $\mathcal H$. For instance, we are interested only in systems with translational invariance  and $\mathbb Z_2$ symmetry. These requirements  imply that the only terms allowed in $\ham$ must contain an even number of spins and their interaction depends only on the distance between spins. An example of such an Hamiltonian is
\begin{equation}
	- \ham[\bs] = \frac 12  \sum_{\bn,  \boldsymbol m } K_{|\bn- \boldsymbol{m}|} \sigma_{\bn}\sigma_{\boldsymbol m},
	\label{lamello}
\end{equation}
where  $K_d$ is the coupling constant between  spins separated by a distance $d$. The form in Eq.~\eqref{lamello} is just an anstaz because  there are many other terms that could have been included, but for    simplicity, we will  consider only Hamiltonians of the form in Eq.~\eqref{lamello}.  

We will   simplify the problem even further. In a system of linear dimension $L$ the maximum distance between two spins is $d = L$, and,  in principle, we should take into account $L$ possible couplings $\{\kk_1, \dots, \kk_L\}$. We will  assume that the couplings are zero greater than a certain distance $d_{\max}$. In all the numerical results shown in the following we use $d_{\max} = 4$.

Finding the Hamiltonian of the form in Eq.~\eqref{lamello} that is most likely to generate the data in Eq.~\eqref{datta} is a well posed question.  Given the couplings $\kkk =\{\kk_1, \, \kk_2, \dots, \kk_{d_{\max}} \}$,  the probability of observing a given configuration $\bs$ is,
\begin{align}
\mathbb P(\bs | \kkk) & = \frac 1 {Z(\kkk)} \exp \left (\frac 12  \sum_{\bn, \, \boldsymbol m }  {K} _{|\bn- \boldsymbol{m}|} \sigma_{\bn}\sigma_{\boldsymbol m} \right)\qquad, 
	\label{likelihood_P}	\\ 
Z(\kkk) & = \sum_{\bs}\exp \left (\frac 12  \sum_{\bn, \, \boldsymbol m }  {K} _{|\bn- \boldsymbol{m}|} \sigma_{\bn}\sigma_{\boldsymbol m} \right).
	\label{likelihood_partition}
\end{align}
Therefore,  the probability of observing the data in Eq.~\eqref{datta} is, 
\begin{equation}
	\mathbb P (\mathrm{data} | \kkk) = \prod_{t =1}^{N_s}\frac 1 Z  \exp \left (\frac 12  \sum_{\bn, \, \boldsymbol m }  {K} _{|\bn- \boldsymbol{m}|} \sigma_{\bn}(t)\sigma_{\boldsymbol m}(t) \right).
\end{equation}
We use Bayes' theorem, or  simply the property of conditional probabilities $p(x|y)p(y) = p(y|x)p(x)$, and obtain
\begin{equation}
	\mathbb P(\kkk| \mathrm{data}) = \frac{\mathbb P(\mathrm{data}| \kkk) \mathbb P (\kkk)}{\mathbb P (\mathrm {data})} ,
	\label{likelihood}
\end{equation}
where $\mathbb P (\mathrm{data})$ is the probability of the data averaged over all possible parameters $\kkk$; $\mathbb P (\kkk)$ is the  a priori  probability of the parameters $\kkk$, and for this reason is called \textit{prior}.  The quantity in Eq.~\eqref{likelihood} is  the \textit{likelihood} and  represents the probability that the data in Eq.~\eqref{datta} have been generated by a model of the type in Eq.~\eqref{lamello} with parameters $\kkk$.

The maximum likelihood approach consists of finding the value of the couplings $\kkk^\star$ that maximizes the likelihood,
\begin{equation}
		\kkk^\star =\max_{\kkk}    \mathbb  P (\kkk | \bs)  .
\end{equation}

If we assume that the prior $\mathbb P(\kkk)$ is constant, and because the probability of the data $\mathbb P (\mathrm{data})$ is independent of $\kkk$, the likelihood in Eq.~\eqref{likelihood}  takes the  form,  
\begin{equation}
 \mathbb P (\kkk | \mathrm{data}) = \mathbb P_0  \prod_{t =1}^{N_s}\frac 1 Z  \exp \left (\frac 12  \sum_{\bn, \, \boldsymbol m }  {K} _{|\bn- \boldsymbol{m}|} \sigma_{\bn}(t)\sigma_{\boldsymbol m}(t) \right),
\end{equation}
where $\mathbb P_0$ is a constant independent of $\kkk$. It is useful to define the logarithm of the likelihood $\mathcal L(\kkk)  = \frac 1 {N_s}\ln(\kkk | \mathrm{data})$, which up to a constant term is
\begin{equation}
	\mathcal L(\kkk)  =  
	\frac 12  \frac 1 {N_s} \sum_{t = 1} ^{N_s} \sum_{\bn, \boldsymbol{m}} {K} _{|\bn- \boldsymbol{m}|} \sigma_{\bn}(t)\sigma_{\boldsymbol m}(t)
	-\ln(Z) .
	\label{logL}
\end{equation}
The solution of the inverse Ising problem is given by the  value of ${\kkk}$ that maximizes $\mathcal L(\kkk)$.

\subsection{The solution of the maximum likelihood problem} 

The solution $\kkk^\star$ of the maximum likelihood problem is the stationary points of the likelihood and satisfies
\begin{equation}
	\frac{\partial \mathcal L(\kkk)}{\partial \kk_d} =0 \qquad    (d = 1, \dots, d_{\max}),
	\label{ML0}
\end{equation}
where the derivative of the log-likelihood is given by
\begin{equation}
	\frac{\partial \mathcal L(\kkk)}{\partial  \kk_d} = 2 d  \left [ C_{\mathrm{data}}(d) - C_\mathbb P (d; \kkk) \right].
	\label{ML}
\end{equation}
$C_{\mathbb P }(d; \kkk)$ is the correlation function at distance $d$ computed using the probability distribution in Eq.~\eqref{likelihood_P} and $C_{\mathrm{data}}(d)$ is the correlation function computed from the data, 
\begin{align}
C_{\mathbb P} (d; \kkk) &	 =  \frac 1 Z \sum_{\bs\}}  
\frac 1{4 d} \sum_{|\bn - \boldsymbol{m}| = d} 
\sigma_{\bn } \sigma_{\boldsymbol {m}}   \, e^{\frac 12  \sum_{\bn, \, \boldsymbol m }  {K} _{|\bn- \boldsymbol{m}|} \sigma_{\bn}\sigma_{\boldsymbol m}} \\ 
C_{ \mathrm{data}} (d) &	 = \frac{1}{N_s} \sum_{t = 1}^{N_s}
\frac 1{4 d} \sum_{|\bn - \boldsymbol{m}| = d} 
\sigma_{\bn} (t) \sigma_{\boldsymbol {m}}  (t).
\end{align}

The maximum of the likelihood can be found iteratively using the gradient descent technique, which is a standard method of numerically finding the maximum/minimum of a function. We start from a random value of the couplings $\kkk$   and iteratively update the couplings according to the rule, 
\begin{equation}
	\kkk^\prime = \kkk + \eta \frac{\partial \mathcal L (\kkk)}{\partial \kkk}.
	\label{ML_iteration} 
\end{equation}
The gradient descent technique might fail if the function we are maximizing has many local maxima. In this case, the solution we find depends on the random initialization, and there is no guarantee that we will find the global maximum of the function. In our  case, it can be shown that the likelihood $\mathcal L$ is a strictly concave function with a unique  maximum.\cite{Zecchina}  Therefore, if we iterate  Eq.~\eqref{ML_iteration}, the couplings $\kkk$ converge to the maximum likelihood solution Eq.~\eqref{ML0}.\cite{Zecchina} The parameter  $\eta$ is called the learning rate, and   determines how fast we approach the solution $\partial_{\kkk} \mathcal L = 0$. Ideally, we would like to choose $\eta$ as high as possible, but  if  $\eta$ is too large, the gradient descent algorithm might become unstable. 

We still have a problem finding the maximum because we have to iterate Eq.~\eqref{ML_iteration} many times   for $\kkk$ to converge, and at each iteration we have to compute $C_{\mathbb P}$, meaning that we have to sample from the probability distribution Eq.~\eqref{likelihood_P}. Computing moments of Eq.~\eqref{likelihood_P} requires using Monte Carlo methods. It is evident that this approach is computationally  costly. 

\medskip \noindent \textit{Problem 4:  The maximum likelihood equation}.
Derive  Eq.~\eqref{ML} by  explicitly taking the derivative of the log-likelihood Eq.~\eqref{logL} with respect to $ K _d$.

\subsection{Maximum pseudo-likelihood} 
The algorithm we now describe is an adaptation of Auerll and Ekeberg.\cite{IIP_uald} 
Again we compute  the set $\boldsymbol \sigma$ of $N_S$ independent samples of a 2D Ising model of linear dimension $L$.
Our goal is to find a model of the form Eq.~\eqref{lamello} that describes the experimental data   in Eq.~\eqref{datta}. For each of the variables $\sigma_{\bn}$ we consider the conditional probability of $\sigma_{\bn}$ given all the other variables. This probability is given by, 
\begin{equation}
	\mathcal P (\sigma_{\bn}|\bs_{\backslash \bn}; \kkk) = 
	\frac{1}{ 1+ e^{-2 \sigma_{\bn} H_n}     } \qquad \quad 	H_{\bn} = \sum_{\boldsymbol{ m}}   K_{|\bn - \boldsymbol{m} |} \sigma_{\boldsymbol{m}}, 
	\label{PabstBlueRibbon} 
\end{equation}
where $H_n$ is the local field acting on  spin $\sigma_{\bn}$, and $\boldsymbol \sigma_{\backslash \bn}$ is the set of all spins except the spin at site $\bn$. If we consider $\sigma_{\bn}$ as the dependent variable, and the complementary set $\sigma_{\backslash \bn}$ is taken as independent variables, the maximum likelihood estimator for $\kkk$  is
\begin{equation}
 \ell_{\bn} (\kkk) =  \frac1 {N_s} \sum_ {t = 1} ^{N_s} \ln \mathcal P(\sigma_{\bn }^t|\boldsymbol \sigma^t _ {\backslash  \bn  }; \kkk),
\label{NarraGunsett} 
\end{equation}
where  $\sigma_{\bn}^t = \sigma_{\bn}(t)$.
We define the pseudo-likelihood $\mathcal L _P (\kkk)$ of the system as 
\begin{equation}
	 \mathcal L_{P}(\kkk)  = \sum_{\bn} \ell_{\bn}(\kkk) .
	\label{Peroni} 
\end{equation} 

Similar to what we described for the maximum likelihood approach, the solution of the inverse Ising problem is given by the maximum of the pseudo-likelihood, 
\begin{align}
	\kkk ^\star & = \mathrm{argmax}_{\kkk} \mathcal L_P(\kkk) \\
	\frac{\partial  \mathcal L_P (\boldsymbol{\kkk},\bs)}{\partial \kk _ d} &  = 
	 \frac{2}{L^2 N_s} \sum_{t = 1}^{N_s} \sum_{\bn}
	 \sum_{|\bn- \bmm|=d}
	 \frac{\sigma_{\bn}(t) \sigma_{\bmm}(t)}{1+e^{-2 \sigma_{\bn}(t)  \sum_{\boldsymbol l } \kk_{|\boldsymbol l  -\bn|}  \sigma_{\boldsymbol l }(t)  }}  .
	\label{Heineken}
\end{align}

As we did to  find the solution of the maximum likelihood problem, we can find the solution to the pseudo-likelihood problem using a gradient descent algorithm. Hence, we iterate the  equation, 
\begin{equation} 
	\kk_d ^\prime = \kk _d   + \eta \frac{\partial  \mathcal L _P  (\boldsymbol{\kkk},\bs)}{\partial  \kk _ d}  \qquad  \qquad     (d = 1, \dots, d_{\max})   . 
	\label{StacysMom} 
\end{equation} 
However, in this case we can compute the gradient of the likelihood directly from the data using  Eq.~\eqref{Heineken}.

Equation~\eqref{StacysMom} implements the simplest possible optimization scheme and   is  sufficient to solve our simple problem. For more complex problems more sophisticated techniques are usually needed.\cite{Zecchina}  Surprisingly, even if the log pseudo likelihood~\eqref{Peroni} is just an approximation of the  log likelihood~\eqref{logL}, it can be proven that in the large sample size limit (see Problem 5) the pseudo-likelihood maximization method gives the correct solution.

\begin{figure}[h]
	\centering
	\includegraphics[width = 0.75\textwidth]{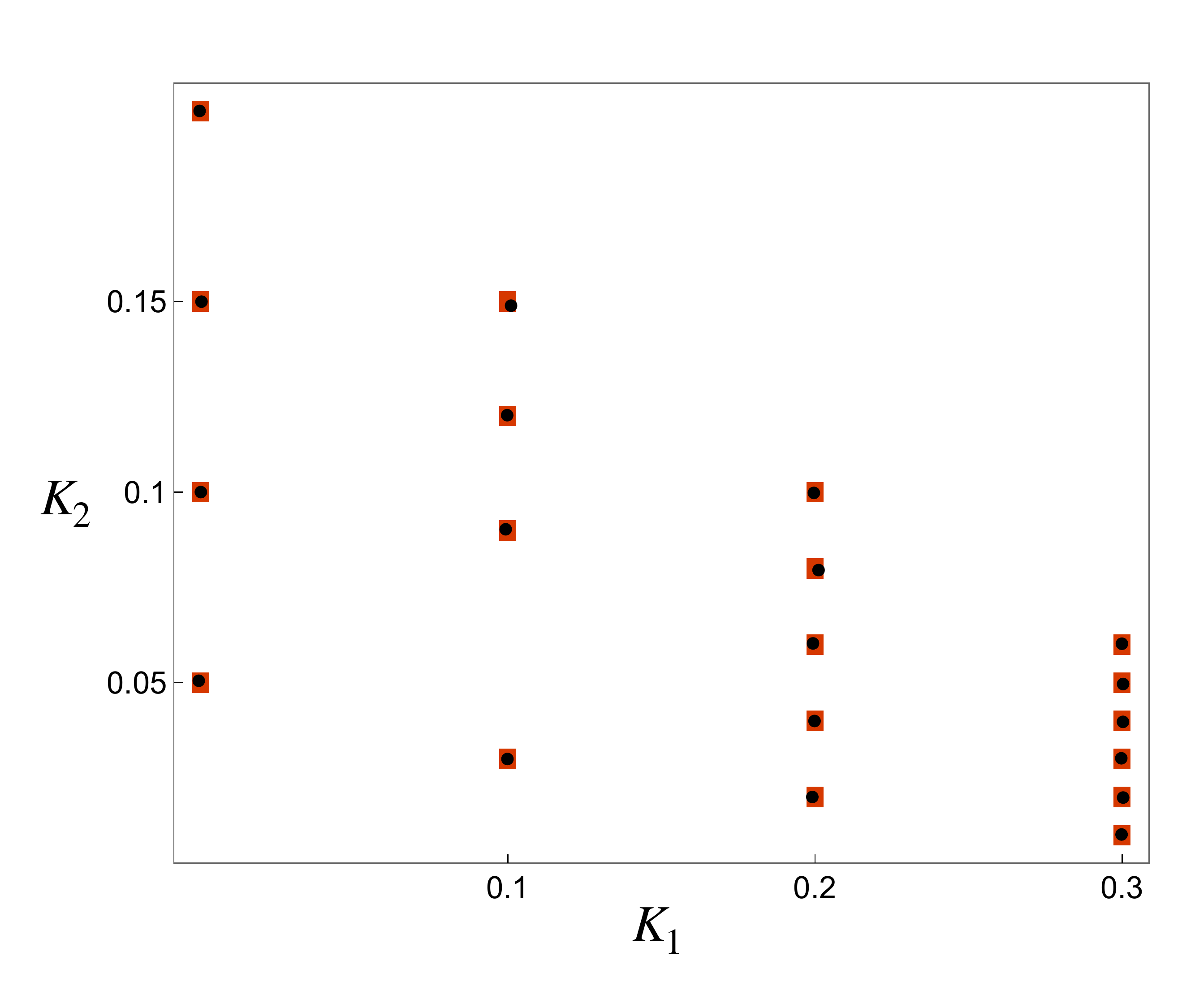}
	\caption{Testing the pseudo-likelihood maximization inference method.   The red squares are  the values of the couplings $(\kk_1,\kk_2)$ used to generate the samples,  the black dots correspond to the parameters retrieved by the configurations, obtained by maximizing the pseudo-likelihood $\mathcal L_P$ in Eq.~\eqref{Peroni} using $d_{\max} = 4$. The data are from a  simulation of the 2D Ising model  with $L = 120$ and periodic boundary conditions. The reader should be able to reproduce these results using the ``Inference-Inverse Ising Problem" section of the MeasureRGflow.ipynb python notebook in Ref.~\onlinecite{suppl}.}
	\label{BlueMoon}
\end{figure}

We can check that the pseudo-likelihood inference scheme works as expected. We generate independent equilibrium configurations from an Ising model with just nearest-neighbor $\kk_1$ and next-nearest neighbor $\kk_2$ interactions, and check whether or not the pseudo-likelihood inference method can correctly retrieve the couplings from the configurations.  The results are shown in Fig.~\ref{BlueMoon}.

\medskip \noindent \textit{ Problem 5: Correctness of the pseudo-likelihood approach}. 
Assume that you have a very large sample, $\{ \sigma_{\bn} (t)\}_{\bn = 1, \dots, L^2} ^{t = 1, \dots, N_s}$, $N_s \to \infty$, extracted from a probability distribution of the form Eq.~\eqref{likelihood_P} and with couplings $\tilde \kkk$.  Show that $\mathcal L_{P}(\kkk)$  in Eq.~\eqref{Peroni} is a maximum at $\kkk = \tilde \kkk$ (Hint: show that in this case the derivative of the pseudo-likelihood in Eq.~\eqref{Heineken} is zero at $\kkk = \tilde \kkk$).  

\medskip \noindent \textit{ Problem 6: Testing the pseudo-likelihood approach}.
Apply the pseudo-likelihood maximization approach, { with $\eta = 0.02$}, to the data generated in Problems~1 and 3 to retrieve the coupling constants used in running the simulations.

\section{Measuring the renormalization group flow} 
\label{numerical}

Our goal is to measure numerically the renormalization group flow, given that  we have all the necessary tools. The strategy we will use is  straightforward. We sample independent equilibrium configurations,   Eq.~\eqref{configurations}, from a 2D Ising model with nearest neighbor $\kk _1$ and next-nearest neighbor $ \kk_2$ interactions with periodic boundary conditions. Then we coarse grain the configurations of the system according to the Kadanoff blocking rule in Eq.~\eqref{Keglevich}; that is, we group variables in blocks of size $b$ and obtain the blocked configurations   $\boldsymbol \sigma^{(b)}(t)$
\begin{equation}
	\boldsymbol \sigma^{(b)} = \{ \sigma_i(t)\}_{i =1\dots,(L/b)^2} ^{t = 1,\dots, N_s}.
	\label{balocco}
\end{equation}
Note  that in a finite-size system the coarse-graining procedure reduces the linear size of the system by a factor of $b$.

\begin{figure}[h]
	\centering
\includegraphics[width = \textwidth]{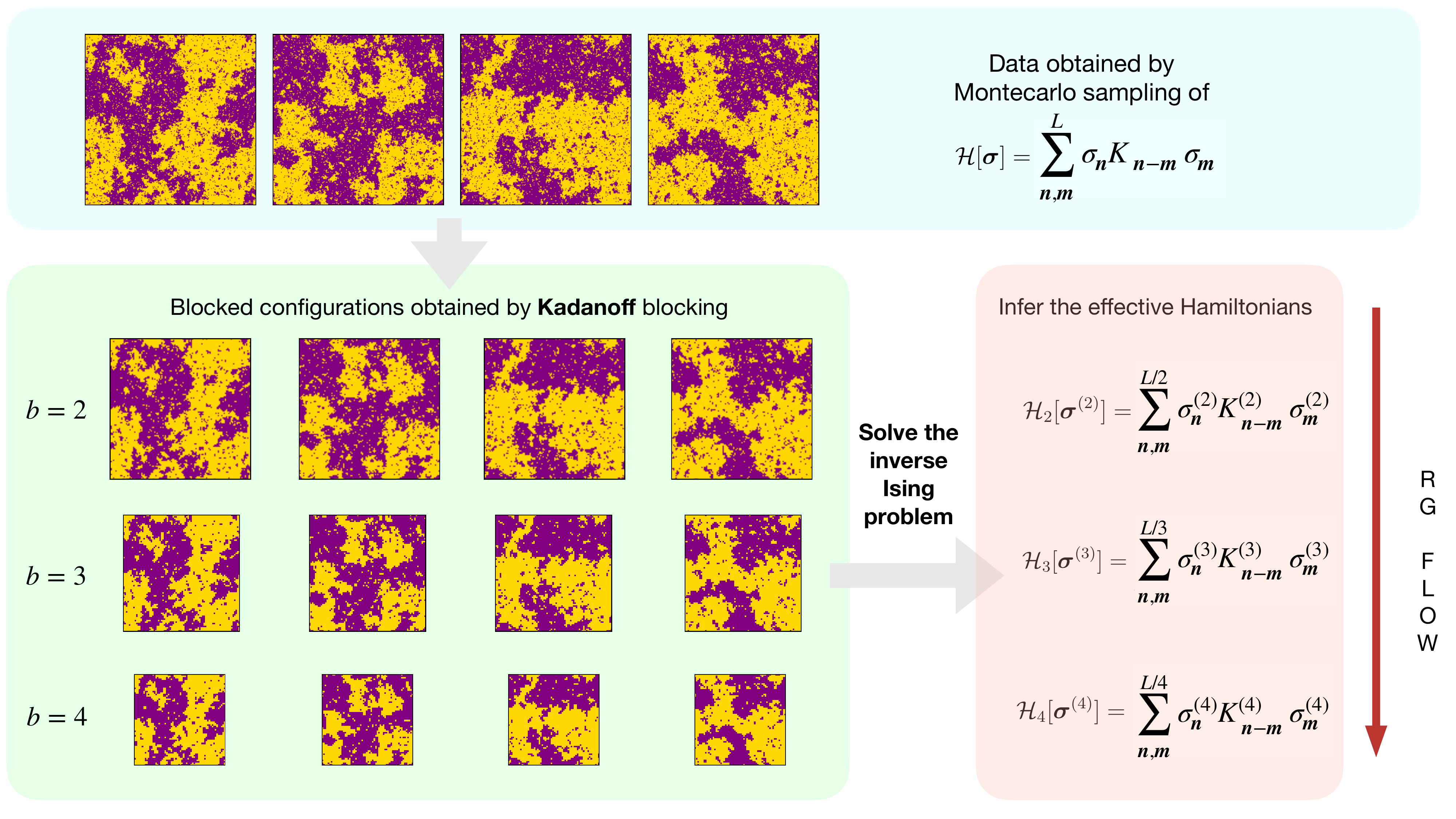}
\caption{Summary of the numerical method used to measure the renormalization group flow. We sample  equilibrium configurations from an equilibrium Ising model (blue) and  apply the Kadanoff blocking rule in Eq.~\eqref{Keglevich} for increasing values of the block size $b$ to obtain a coarser and coarser description of the system in which small scale fluctuations are integrated out (green). Finally, we apply the inverse Ising procedure to infer the Hamiltonian that would have generated the coarse grained data (red). The result of this procedure is the renormalization group flow of the system, which is how the effective theory changes at different levels of coarse-graining.  }
\label{SummaryExperiment} 
\end{figure}

At this point, we can apply the inverse Ising method described in  Sec.~\ref{IIP} to infer the parameters that would have generated the coarse-grained data in Eq.~\eqref{balocco}. The result of the inference is the effective theory at the $b$th level of coarse-graining. We repeat this procedure for increasing values of the coarse-graining block size. In this way, we  measure how the parameters of the model change under coarse-graining; that is, we are measuring the renormalization group flow. This procedure is summarized in Fig.~\ref{SummaryExperiment}.

We usually think of the renormalization group as an iterative procedure. We iterate the Kadanoff blocking transformation, and after each iteration, we obtain a coarser description of the system. If the linear dimension of the system is $L$, and we iterate the blocking transformation with $b = 2$,  we  obtain coarse grained systems of sizes $L/2$, $L/4$, $L/8$, and so on. If we compare the coarse grained systems to the original one, this procedure  corresponds to block sizes $b = 2,  4,  8, \dots$. 
The main  downside of this approach is that the size of the system decreases exponentially with the number of renormalization group steps (the number of Kadanoff blocking iterations). 
An alternative is to apply the blocking transformation to the original system for increasing values of the block size 
$b = 2,  3,  4,  5,   6, \ldots$.  In this  way we obtain blocked systems of sizes $L/2$, $L/3$,  $L/4, \ldots$ . By using this  strategy the size of the system  decreases linearly with the number of steps, and for this reason we  adopt the latter strategy.

\begin{figure}[h]
	\centering
	\includegraphics[width = 1\textwidth]{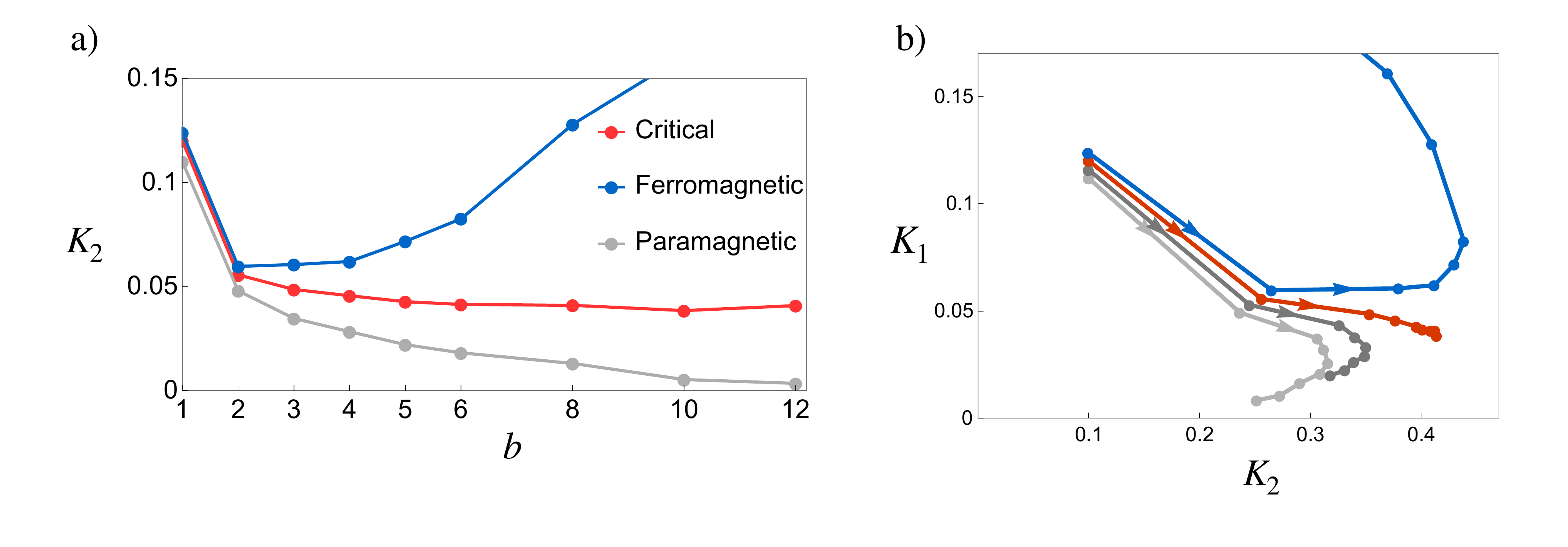}
	\caption{Numerical renormalization group flow.
(a) Flow of the next-nearest neighbor coupling $\kk_2$ as a function of the block size $b$, with $\kk_1 = 0.1$ for different initial values of $\kk_2$. Depending on the value of  $\kk_2$,  we observe different qualitative behavior of $\kk_2^{(b)}$ for large $b$: $\kk_2^{(b)} \to 0$ paramagnetic  phase (gray curve), $\kk_2^{(b)} \to \infty$ ferromagnetic phase (blue curve), and $\kk_2^{(b)} \to \kk^* _2$ critical regime (red curve). (b) Renormalization group flow projected  on the $(\kk_1,\kk_2)$ plane; the arrow goes toward increasing values of    $b$. Different colors correspond to different regions of parameter space: paramagnetic (gray), ferromagnetic (blue), and critical (red). The numerical renormalization group flow is compatible with the qualitative picture of Fig.~\ref{seborrea}. Data are for an initial value of  $L = 480$ and $N_s = 3\times 10^3$. The reader should be able to reproduce  these results following the MeasureRGflow.ipynb python notebook in Ref.~\onlinecite{suppl}.}
\label{borra}
\end{figure}

We first simulate a system of size $L = 480$ for different values of the couplings $(\kk_1,\kk_2)$. In particular, we fix    $\kk_1 = 0.1$ and simulate systems with different values of $\kk_2$. For each of these simulations, we perform the blocking transformation with block sizes $b = 2,3,4,5,8,10,12$ and, following the procedure described in Fig.~\ref{SummaryExperiment}, we measure the resulting renormalization group flow. The results are summarized in Fig.~\ref{borra}. We first  note that the  simulations reproduce correctly the qualitative  picture illustrated in Fig.~\ref{seborrea}. We can clearly distinguish the  different qualitative behavior depending on the phase of the system. In the paramagnetic phase (gray curves) the  flow converges to infinite temperature or zero coupling theory. In the ferromagnetic phase (blue curve) the  flow converges to a zero temperature (strong coupling) theory. Finally, as predicted by the  theory, in the critical regime (red curve) the   flow converges to a finite fixed point. This behavior implies that in the critical regime the renormalization group flow converges to a finite fixed point.

\begin{figure}[h]
	\centering
	\includegraphics[width = 0.9 \textwidth]{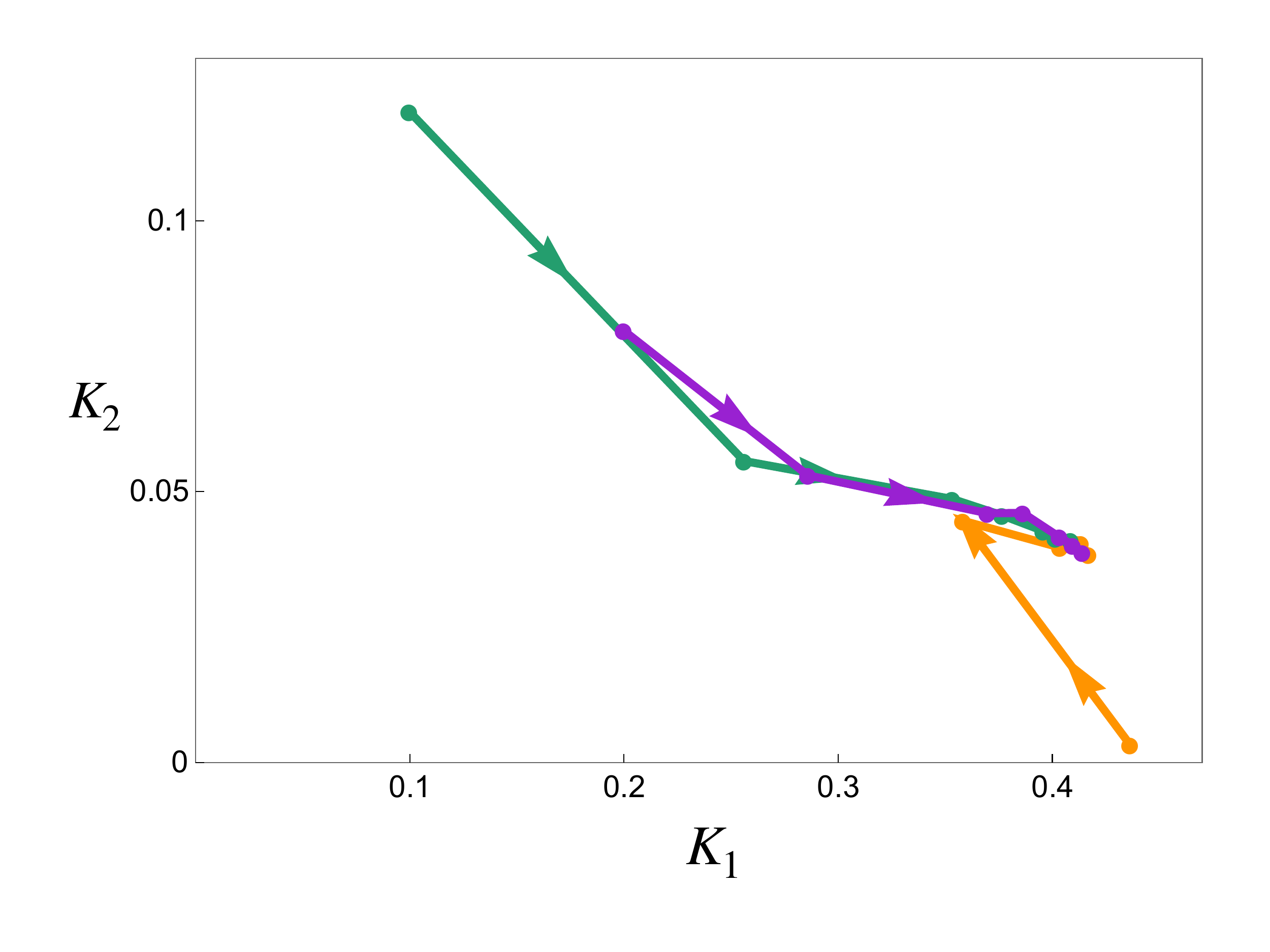}
	\caption{The renormalization group on the critical manifold. Three   trajectories, relative to $(\kk_1,\kk_2)$ on the critical manifold. Any renormalization group flow trajectory that starts on the critical manifold converges to the same fixed point, independently of the initial value of the pair  $(\kk_1,\kk_2)$. Data for $L = 480$ and $N_s = 3\times 10^3$.}
	\label{ccamado}
\end{figure}

We still need to show that this fixed point is unique and that any renormalization group flow trajectory starting on the critical manifold converges to the same fixed point. Different pairs $(\kk_1, \, \kk_2)$ can correspond to a critical Ising model, with a divergent correlation length $\xi$, as shown in Fig.~\ref{amazza}. In this  numerical experiment we simulate the Ising model with three  pairs of couplings $(\kk_1, \, \kk_2)$, carefully tuned to be on the critical manifold. The prediction of  renormalization group theory is that the flow converges to the same  fixed point. As   before, we apply the blocking transformation with block sizes $b = 2,3,4,5,8,10,12$ and determine the resulting renormalization group flow.  The results  are shown in Fig.~\ref{ccamado}. In  agreement with the theoretical picture in Eq.~\eqref{seborrea},  the three trajectories converge to the same fixed point. 
The yellow curve starts from $(\kk_1=0.44,\, \kk_2 = 0)$, meaning that at the beginning of the renormalization group flow the system does not have any next-nearest neighbor interactions, $\kk_2 = 0$.  The next-nearest neighbor interaction is generated by the coarse graining procedure, which is an example of an interaction term that is  generated by the renormalization group. Also,  interactions at larger distances are generated such as $\kk_3$ and $\kk_4$. In the numerical experiment showed in Fig.~\ref{ccamado} the fixed point is at $\kk_3 = 0.011$ and $\kk_4 = 0.031$. {Interestingly the fixed point values of the higher distance couplings are significantly different form zero. The dominant term is the expansion is the nearest neighbors coupling $K_1$, which is an order of magnitude larger.   }

The  numerical experiment we have presented is restricted to the     Ising model. However,  renormalization group theory predicts that a similar mechanism explains universality in more realistic systems, and that the very same fixed point theory shown in Fig.~\ref{ccamado} also describes  the critical properties of some real ferromagnets.\cite{Cardy}

\section{Conclusion} 
By using a combination of Markov-chain Monte Carlo, Kadanoff blocking, and  inverse Ising model calculations, we were able to study  the renormalization group flow in the ferromagnetic Ising model. The results are in   agreement with    renormalization group theory.\cite{Cardy} The numerical experiments could be helpful for consolidating the reader's understanding of  renormalization group theory.  Although  renormalization group theory has been tested in several contexts, to the best of my knowledge, this is the first time that this particular aspect of the renormalization group has been studied numerically, which  may be attributed to the fact that, during the early stages of the development of renormalization group theory, the inverse Ising model techniques were not yet developed and the computational  power was insufficient for such a numerical experiment. It is fascinating how the numerical results match perfectly with the theoretical predictions that were conceived almost fifty years ago.
 
\section*{acknowledgments}
L.D.C.  is supported by  the CPBF fellowship (NSF PHY-1734030). I  thank F. Ferretti, R. Pang and F. G. Castro for comments upon reading the manuscript. I thank A. Cavagna for fruitful discussions and U. Tauber for suggesting the publication of this manuscript.


\begin{thebibliography}{99}
\bibitem{MCMC_is_hard}Computing averages over the Ising probability distribution  in Eq.~\eqref{BasicIsingP} is not  trivial. The reason is that the number of possible configurations $C=2^{L^2}$ of the system is exponentially large with the number of spins. It is possible to numerically sample the probability distribution    using the Markov-chain Monte Carlo technique.\cite{MCMC,MCMC2}
\bibitem{Escaping}How trajectories that are close to the critical regime run away from the critical manifold determines the scaling properties of the system. We will not focus on this particular aspect; more details can be found in standard textbooks such as Chaps.~3   of Ref.~\onlinecite{Cardy} and  Chap.~4 of Ref.~\onlinecite{Kardar}.
\end{thebibliography}
\end{document}